\documentclass[fleqn,12pt,twoside]{article}
\usepackage{espcrc1}

\usepackage{graphicx}
\usepackage[figuresright]{rotating}

\newcommand{\AmS}{{\protect\the\textfont2
  A\kern-.1667em\lower.5ex\hbox{M}\kern-.125emS}}

\def\pislash{ {\pi\hskip-0.6em /} }
\def\nopi{ {\rm EFT}(\pislash) }
\def\si{{}^1\kern-.14em S_0}
\def\siii{{}^3\kern-.14em S_1}
\def\diii{{}^3\kern-.14em D_1}

\title{Effective Field Theory for Nuclear Physics}

\author{Martin J. Savage\address[UW]{Department of Physics, 
University of Washington,
Seattle, WA 98195-1560, USA. }
       }
       
\begin{document}

\maketitle

\begin{abstract}
I review the current status of the application of effective field theory to
nuclear physics, and its present implications for nuclear astrophysics.
\end{abstract}

\section{Introduction}

When we finally arrive at the complete effective field theory
that faithfully reproduces QCD in the low-energy regime relevant for nuclear
physics,
we will be able to answer some fundamental questions.
In addition, to being able to describe inelastic processes,
such as $3\alpha\rightarrow \ ^{12}{\rm C}+\gamma$,
in the same framework as elastic processes and the framework with which the
nuclear energy levels are computed to high accuracy, we will know how such
processes depend upon the fundamental constants of nature, 
the strong interaction mass-scale, $\Lambda_{\rm QCD}$, and 
the masses of the quarks, $m_q$.
A small but significant  step has been taken in this direction by a recent
calculation of the behavior of the two nucleon systems as a function
of $m_q$.
There are three good reasons for understanding such behavior.
The first reason is  intellectual curiosity. 
If we believe that we completely understand the strong interaction we should 
be in a position to address this issue.
The second reason is to put constraints on physics beyond the standard model.
There are hints that the constants of the standard
model, such as $\alpha_{\rm em}$ and $m_q$, 
may be time-dependent~\cite{Webb}.
The third and perhaps the most practical reason is that 
the $m_q$-dependence of nuclear properties and reaction rates 
needs to be known 
in order to extrapolate lattice-QCD  calculations 
from the unphysical values that will be used in present simulations
and in those of the near future,
down to those of nature.
For the foreseeable future lattice QCD calculations will be
performed with unphysically large quark masses simply due to the fact that 
the time required to perform the
simulations increases as one reduces the $m_q$.
Therefore, effective field theory (EFT) will play 
a key role in any future lattice
QCD program, as it will only be through matching onto the appropriate field
theory that lattice QCD will be able to make predictions of physical
observables.
This is true, of course, only until such a time when
such observables can be computed directly with the physical
values of the quark masses.
The link between EFT and data has been and continues to be very
strong, with data constraining the finite number of counterterms that appear at
any given order in the EFT expansion.
However, EFT calculations are being performed at sufficiently high orders so
that in some cases the number of counterterms is greater than the number of
observables, and lattice-QCD will be the only way to further improve the
calculation.
At this time constraints are being placed on linear combinations of the
Gasser-Leutwyler coefficients, the $L_i$, through partially-quenched QCD
(PQQCD) simulations~\cite{pqqcdLi}.

For processes involving multiple nucleons it is convenient to describe
different momentum regimes with different EFT's.
If one is dealing with processes in which all momenta are less than the
pion mass, $m_\pi$, it proves to be useful to use the pionless EFT, $\nopi$,
while for processes involving momenta greater than $m_\pi$
one must use a pionful theory.

\section{The $p \ll m_\pi$ Regime}
In the kinematic regime where $|{\bf p}|\ll m_\pi$ we can construct an EFT
to describe the interactions of multiple nucleons and photons quite
simply~\cite{nopi,KSW98,ch99}.
The fact that there is a bound state near threshold in the $\siii-\diii$
coupled-channels, and a pole on the second-sheet near threshold in the $\si$
channel means that at least one operator in the EFT Lagrange density must be
treated non-perturbatively.   One is free to choose which operators are 
resummed, and this 
will be defined in the power-counting associated with the EFT.
There are no explicit $\pi$'s, $\rho$'s or other hadrons that can be
produced in the low-energy collisions.
As $\nopi$ is only applicable in the very-low energy regime,
chiral symmetry is not a good symmetry, however isospin is a good symmetry.
The only input into the construction of $\nopi$ is Lorentz
invariance, electromagnetic gauge invariance,
baryon number conservation and the approximate isospin
symmetry, the breaking of which is included perturbatively.

The strong interactions between two nucleons are described 
by a Lagrange density
\begin{eqnarray}
{\cal L} & = & C_0  \left[\ N^T P^j N\ \right]^\dagger N^T P^j N
\ +\ ...
\ \ \ ,
\label{eq:strongnopi}
\end{eqnarray}
where $P^j$ is a spin-isospin projector, and the ellipses denote operators
involving more powers of the external energy.
It is clear that there is a correspondence between the operators in 
Eq.~(\ref{eq:strongnopi}) and the coefficients in the effective range (ER)
expansion of the scattering amplitude.
This Lagrange density is explicitly Galilean invariant and relativistic
corrections can be included in perturbation theory 
straightforwardly~\cite{ch99}.
Further, it is straightforward to include the interactions in  higher
angular momentum channels, such as in 
the $\siii-\diii$ coupled channels~\cite{ch99}.
If the only thing that had been accomplished was to recover
effective range theory from $\nopi$, nothing would have been achieved.
However, $\nopi$ is more than effective range theory.
In $\nopi$ one can systematically incorporate gauge fields, such as the photon
and electroweak fields.
In addition to interactions arising from gauging derivatives in the strong
interactions, there are operators that are
gauge-invariant by themselves.
For instance, 
the lowest order (LO)  gauge invariant operator that contributes to the 
deuteron quadrupole moment is
\begin{eqnarray}
  {\cal L}_{\cal Q} & = & -e\ C^{(Q)} 
\left[\ N^T P_i N\ \right]^\dagger \left[\ N^T P_j N\ \right]
\left(\ \nabla^i\nabla^j - {1\over n-1 }\nabla^2\delta^{ij}\ \right) A_0
\ \ \ ,
\label{eq:quadct}
\end{eqnarray}
where 
$A_0$ is the time-component of the electromagnetic field.
A computation of the deuteron quadrupole moment up to next-to-leading order
(NLO)
allows one to fit the coefficient $C^{(Q)}$.
Once this constant is determined other observables 
that receive 
contributions from this $E2$ operator can be computed up to NLO~\cite{ch99}.

\begin{figure}[!ht]
\begin{center}
    \includegraphics*[width=0.4\linewidth,clip=true]{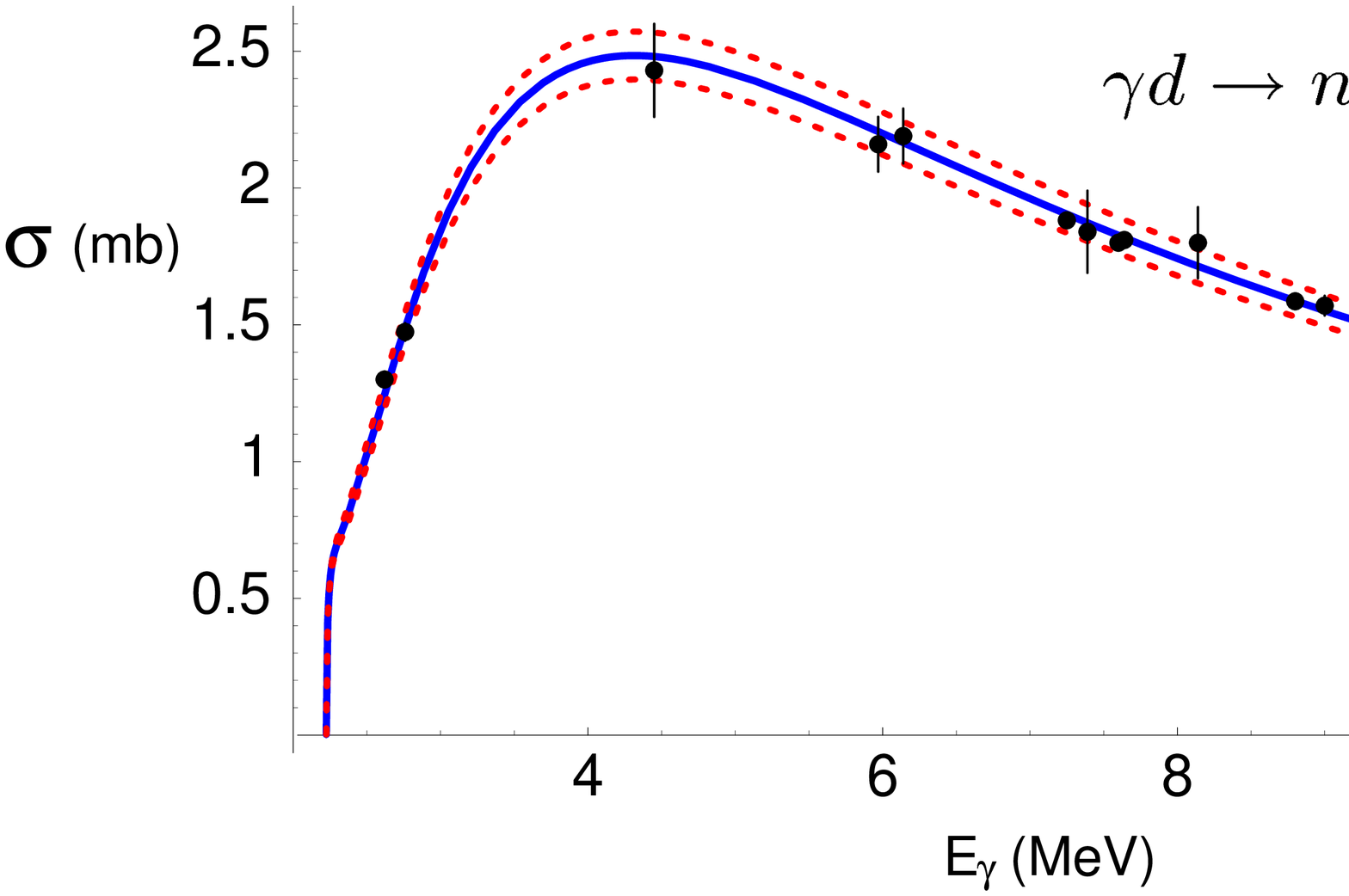}
\hskip 0.6in
    \includegraphics*[width=0.32\linewidth,clip=true]{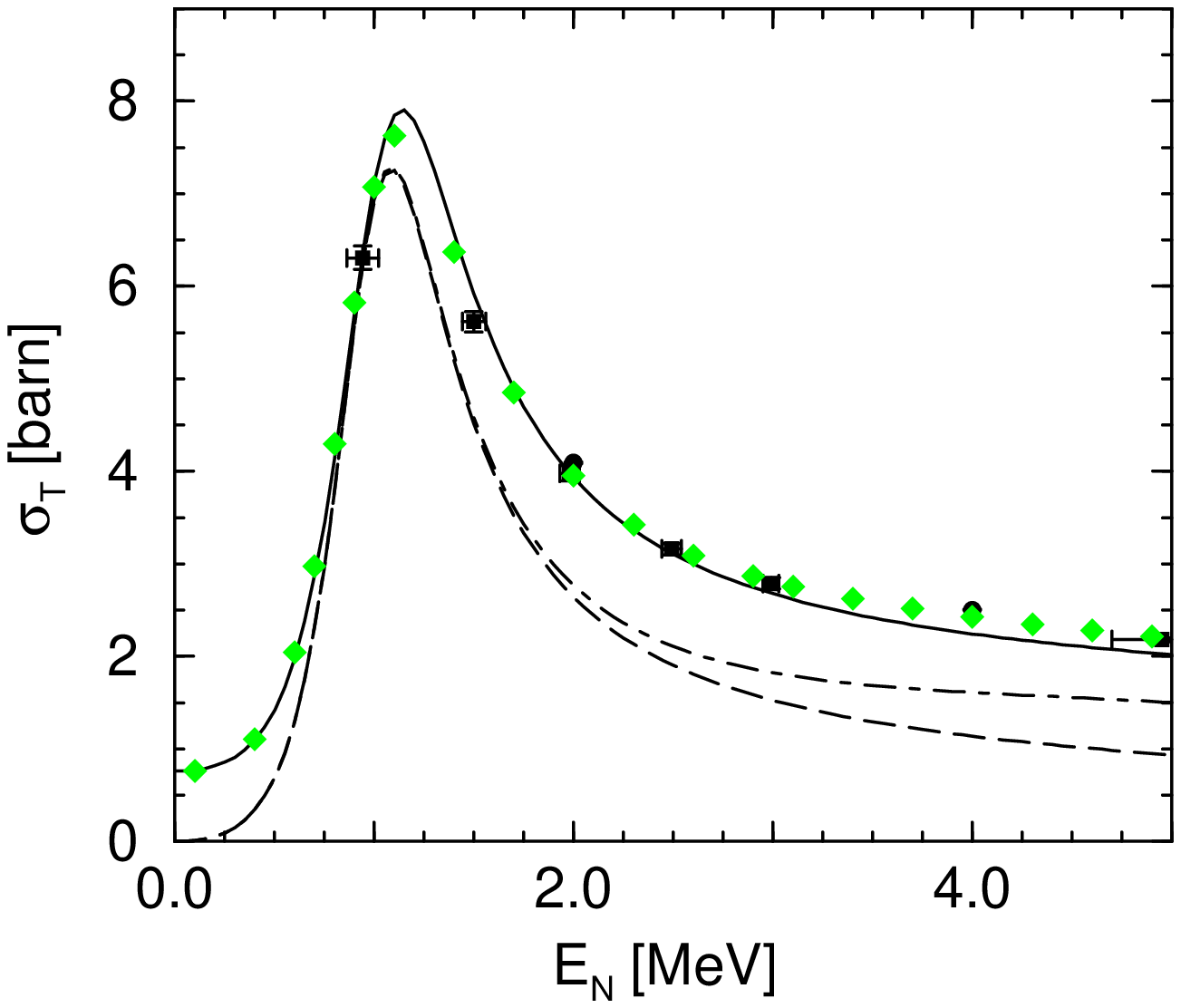}
\end{center}
\vskip -0.5in
\caption{
The left panel shows the cross section for
$\gamma d\rightarrow n p$.
The curves correspond to $L_{np}$ determined by the cross section
for cold $np\rightarrow d\gamma$~\protect\cite{Chen:1999bg} and 
the dashed lines denote the $\sim 3\%$ 
theoretical uncertainty. Rupak has further 
reduced this uncertainty to below $1\%$\protect\cite{Chen:1999bg}.
The right panel shows the total cross section for $n\alpha$ scattering
as a function of the neutron kinetic energy ~\protect\cite{BHvK}. 
The diamonds and black squares are data.
The dashed and solid lines show the EFT results at  LO and NLO, respectively.
}
\label{fig:nopi}
\vskip -0.1in
\end{figure}

The cross
section for $np\rightarrow d\gamma$ at energies relevant to big-bang
nucleosynthesis provides an example of a high precision 
calculation in $\nopi$, as shown in Fig.~\ref{fig:nopi}.
This is a classic nuclear physics process in which meson-exchange
currents play a significant role.
From the $\nopi$ point of view, where  mesons are not an explicit degree
of freedom, such contributions are included via local gauge-invariant
four-nucleon-one-photon interactions
(e.g. with coefficient $L_{np}$ for  $np\rightarrow d\gamma$).
In this case the resonance saturation hypothesis works well, 
in so much as this local operator is saturated by one pion exchange (OPE),
but in general this need not be the case.

Another
place where $\nopi$ has had impact is in the analysis of SNO data,
and the determination of solar neutrino fluxes.
An integral part of the SNO analysis is the cross section for neutrino induced
deuteron break-up, such as the charged current process
$\nu_e d\rightarrow e^- pp$.
In addition to the well-known one-body contributions to the break-up, there
is also  a contribution from two-body interactions.
Butler, Chen and Kong~\cite{Butler:2000zp}
showed that only one constant, $L_{1,A}$,
is required to describe the
two-body contribution in order to calculate the cross-section 
at the percent level.
Therefore, the question is where does one get $L_{1,A}$ from?
So far there have been three independent determinations of $L_{1,A}$:
from the $\beta$-decay of tritium in a hybrid EFT,
from an analysis of reactor data on $\overline{\nu} d$ 
break-up,
and from an analysis of SNO data~\cite{PMSVKRKMR,Lone}.
They find 
$L_{1,A}=6.5\pm 2.4~{\rm fm}^3, 3.6\pm 5.5~{\rm fm}^3, 4.0\pm 6.3~{\rm fm}^3$,
respectively, which are all consistent within errors.

Recently, Bertulani, Hammer and van Kolck~\cite{BHvK} have examined the
relatively simple $n\alpha$ ``Halo-nuclear'' system.  They explored the
cross section for $n\alpha\rightarrow n\alpha$ by treating the $n\alpha$
interaction as a sum of local interactions between a fundamental $n$-field  
and a fundamental $\alpha$-field.
There are three partial waves that make a  significant contributions 
in the low-energy regime, $J^\pi= 0^+, 1^-, 1^+$.
The LO calculation requires the scattering lengths in each channel,
while the NLO calculation requires both the scattering length and effective
range in each.
The experimental values $a_{0^+}=2.4641\pm 37~{\rm fm}$, 
$r_{0^+}=1.385\pm 41~{\rm fm}$,
 $a_{1^-}=-13.821\pm 68~{\rm fm}^3$, 
$r_{1^-}=-0.419\pm 16~{\rm fm}^{-1}$,
 $a_{1^+}=-62.951\pm 3~{\rm fm}^3$, 
and $r_{1^+}=-0.8819\pm 11~{\rm fm}^{-1}$
produce the curves shown in the right panel of Fig.~\ref{fig:nopi}.

\begin{figure}[!htb]
  \begin{center}
    \includegraphics*[width=0.4\linewidth, clip=true]{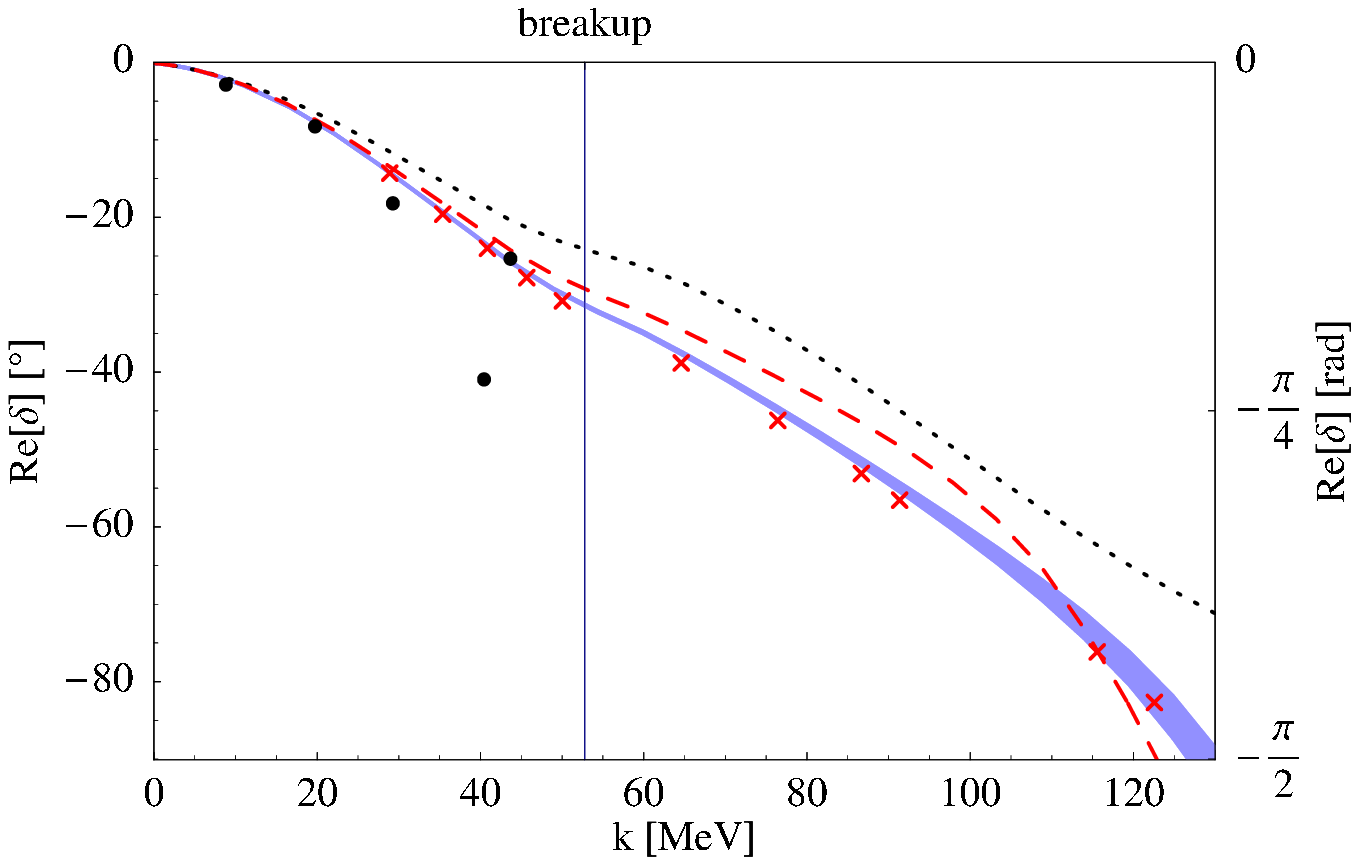}
\hskip 0.6in
    \includegraphics*[width=0.4\linewidth, clip=true]{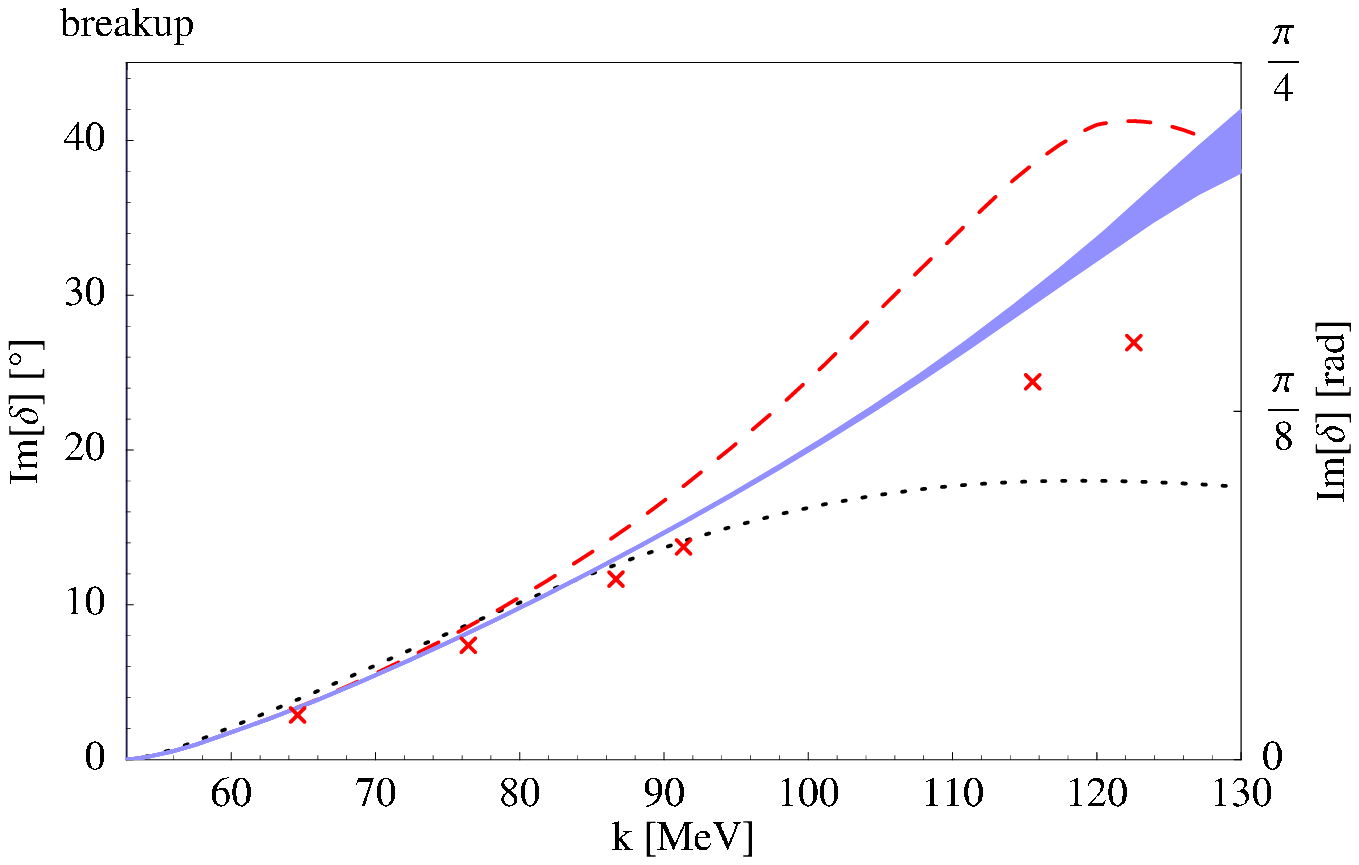}
  \end{center}
\vskip -0.5in
\caption{
  The neutron-deuteron $^2S_\frac{1}{2}$ phase shift 
at LO (dotted line), NLO  (dashed line) and NNLO 
(dark band)~\protect\cite{BGHR}. 
Left: Real part.  Right: Imaginary part.
The band in the NNLO curve results from varying the 
cutoff from $\Lambda=200$ MeV to $\Lambda=600$ MeV. 
The dots are the available direct phase shift analysis of the triton
  channel~\cite{doublet_PSA} and the crosses are the
  results obtained with the Argonne V18$+$Urbana IX two and three body
  forces~\cite{doublet_pots}.}
\label{fig:nd}
\vskip -0.1in
\end{figure}

\begin{figure}[!htb]
  \begin{center}
    \includegraphics*[width=0.4\linewidth,clip=true]{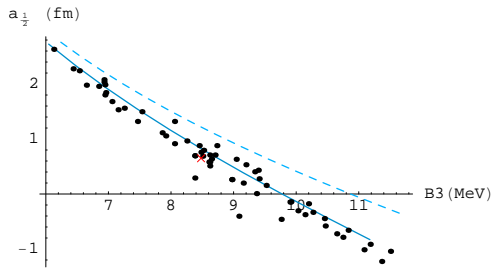}
  \end{center}
\vskip -0.5in
\caption{The Phillips line~\protect\cite{BGHR}. The dots correspond to the
  predictions of different models with the same two body scattering lengths
  and effective ranges~\cite{tkachenko}.  
The dotted and full line are the EFT results at LO
  and NLO, respectively. The cross is nature. }
\label{fig:Phil}
\vskip -0.1in
\end{figure}

There has been continued progress in the three-body sector, 
and I  will discuss only a small fraction of it here.
Bedaque, Grie\ss hammer, Hammer and Rupak~\cite{BGHR} 
have reformulated the EFT 
construction in the triton channel and have produced results for 
the $nd\rightarrow nd$ phase-shift, as shown in Fig.~\ref{fig:nd}.
At LO and NLO there is a momentum independent three-body interaction,
with coefficient $H_0$, that can only be determined in the 
three-body sector (or higher-body sectors), 
and the scattering length in the triton 
channel is sufficient.
At NNLO there is a contribution from a momentum dependent 
three-body interaction, with coefficient $H_2$, and the triton binding energy
is sufficient to determine it.
One sees that the EFT calculation is converging to the data and 
at this order shows no serious deviation from a more  
traditional potential model calculation.
One can also examine the behavior of the Phillips line order-by-order in the
EFT expansion.
If one takes an arbitrary model that describes to arbitrary precision the
two-body sector and uses it to compute the scattering length 
in the $J={1\over 2}$ channel and triton binding
energy, one will generate values that lie (approximately)
on a line in the plane formed by them.
This is the Phillips line, and results from the fact that if three-body
interactions are possible then, in general, they will be present.  
The Phillips line,  as shown in Fig.~\ref{fig:Phil},
is generated by varying  the 
coefficient of the momentum independent three-body force over all possible
values.

There have also been some exciting results from Hammer regarding the 
hyper-triton in $\nopi$~\cite{hyperhammer}.
As the deuteron has only a small binding energy and the hyper-triton has a very
small binding energy, $B_{\Lambda t}= 2.35\pm 0.05~{\rm MeV}$, this
particular three-body system can be described very well in $\nopi$, 
and Hammer obtains 
$a_{\Lambda d}=16.8^{+4.4}_{2.4}~{\rm fm}$ and 
$r_{\Lambda d}=2.3\pm 0.3~{\rm fm}$.
I refer the reader to Ref.~\cite{hyperhammer} for a detailed
discussion.
An improved experimental determination of $B_{\Lambda t}$ would be welcome.

\section{The $p \gg m_\pi$ Regime}

It was Weinbergs' pioneering efforts~\cite{weinberg}
 in the early 1990's that led to the field
of EFT in nuclear physics.
He was attempting to construct an EFT for nuclear processes and nuclei 
involving momentum all the way up to the chiral symmetry breaking scale
$\Lambda_\chi$, and necessarily included the pion as a dynamical degree of
freedom.
The power-counting that he developed, known as Weinberg power-counting (W),
involved performing a chiral expansion of the nucleon-nucleon potential using
the same power-counting rules that apply in the meson and single nucleon
sectors.  The chirally expanded potential is then inserted into the 
Schroedinger equation to determine observables, such as phase shifts.
However, there is a formal problem with this power
counting~\cite{KSW96} 
in some of the scattering channels, particularly the $\si$ channel.
However, extensive phenomenological studies with W power-counting 
appear to be in good agreement with data~\cite{ray,reviews}, 
where such comparisons are possible,
and the formal problem appears to have little impact when a massive
regulator is used with a mass scale that is not radically different from 
a few hundred MeV.
This problem led to KSW power-counting~\cite{KSW98}
in which the momentum
independent four-nucleon operator is promoted to one lower order in the chiral
expansion, and consequently pion exchanges are subleading order and
treated in perturbation theory.
This power-counting is formally consistent and 
gives renormalization group invariant amplitudes
order-by-order in the power-counting.   
However, Fleming, Mehen and Stewart~\cite{FMS} 
(FMS) showed that the scattering amplitude in the 
$\siii-\diii$ coupled channels diverges at NNLO 
at relatively small momenta and KSW power-counting fails.
FMS found that 
a contribution that remains large in the chiral limit destroys
the convergence: it is the chiral limit of the 
tensor force that ``does the damage''.
Recently, it was suggested that one
should expand observables about the chiral limit~\cite{BBSvK}
(BBSvK power-counting).
BBSvK power-counting has all the nice features of W and KSW counting: the 
chiral limit of the tensor
force is resummed at LO along with the momentum and $m_q$ independent
four-nucleon operator in the $\siii-\diii$ coupled channels, while 
pions are perturbative in the $\si$ channel and in higher partial waves where
analytic calculations are possible.

\begin{figure}[!htb]
  \begin{center}
    \includegraphics*[width=0.5\linewidth,clip=true]{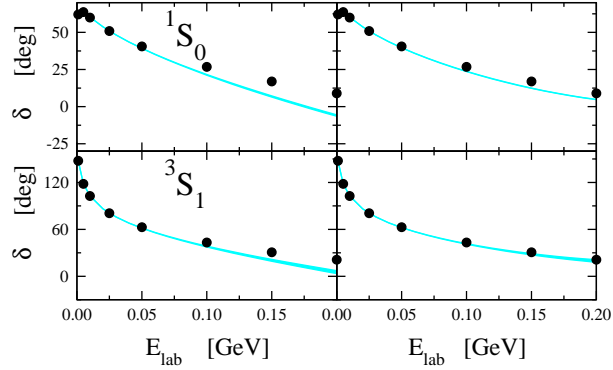}
  \end{center}
\vskip -0.5in
\caption{The S-wave phase-shifts  
for laboratory energies $E_{\rm lab}$ below 
200~MeV~\protect\cite{ENGKMW} . 
Left (right) panel is the result at NLO (NNLO*). 
The momentum-space 
cut-off is chosen between 500 and 600 MeV leading to the band. The
filled circles correspond to the Nijmegen PSA results~\protect\cite{NPSA}. }
\label{fig:Wswave}
\vskip -0.1in
\end{figure}

As I mentioned previously, the phenomenology of W counting has been explored
extensively.
Epelbaum {\it et al}~\cite{ENGKMW} 
have performed an impressive analysis of the
few nucleon systems with W power-counting  and find the calculations 
in the NN sector to converge well.
The S-wave phase shifts are shown in Fig.~\ref{fig:Wswave}.
One sees that higher order calculations move closer to the data in all partial
waves and the uncertainty is reduced.  
This calculation uses a momentum space regulator, $\Lambda$, 
and one finds some sensitivity to the value chosen for $\Lambda$.  
An estimate of the uncertainty 
in this calculation can be made by varying $\Lambda$
between 500 and 600 MeV, however, it would appear that such 
variations in $\Lambda$ tend to underestimate the true uncertainty, as 
the NNLO result does not lie entirely within the NLO band.
The convergence of the theory has also been examined in detail by 
Entem and Machleidt~\cite{EM}.
In other work Epelbaum {\it et al}~\cite{ENGKMW2} 
have performed a detailed study
of the three-body sector with W power-counting.
One finds that the uncertainty is reduced by going from NLO to NNLO as
expected,
and their results are very encouraging.

\begin{figure}[!htb]
  \begin{center}
    \includegraphics*[width=0.4\linewidth,clip=true]{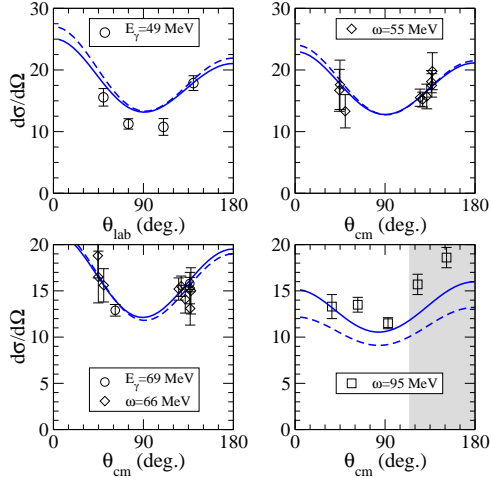}
  \end{center}
\vskip -0.5in
\caption{${d\sigma\over d\Omega}$'s
for Compton scattering on deuterium~\cite{BMMPvK}.
The data are from
Illinois~\cite{lucas} (circles), Lund~\cite{lund} (diamonds) and SAL
(squares)~\cite{SAL}. The solid line is the ${\cal O}(Q^4)$
calculation with $\alpha_N=9.0 \times 10^{-4}~{\rm fm}^3$,
$\beta_N=1.7 \times 10^{-4}~{\rm fm}^3$.  
 The dashed
line is the (parameter-free) ${\cal O}(Q^3)$ calculation.
 }
\label{fig:Compton}
\vskip -0.1in
\end{figure}
Beane {\it et al} have obtained some 
exciting results for Compton scattering from the
deuteron~\cite{BMMPvK}.
By working at order $Q^4$ in W power-counting 
the
electric and magnetic isoscalar polarizabilities of the nucleon 
were determined directly from
the Compton scattering cross section at various energies, as shown in 
Fig.~\ref{fig:Compton}~\cite{BMMPvK}. 
They find that
$\alpha_N=9.0 \times 10^{-4}~{\rm fm}^3$,
$\beta_N=1.7 \times 10^{-4}~{\rm fm}^3$.
A higher order calculation is required in order to reduce the uncertainties in
$\alpha_N$ and $\beta_N$
and to ensure that the EFT calculation of this process is
converging as expected.
Further, it is clear that more data should be taken for this process,
filling out the differential cross sections plots shown in
Fig.~\ref{fig:Compton}.

\begin{figure}[!htb]
  \begin{center}
    \includegraphics*[width=0.4\linewidth,clip=true]{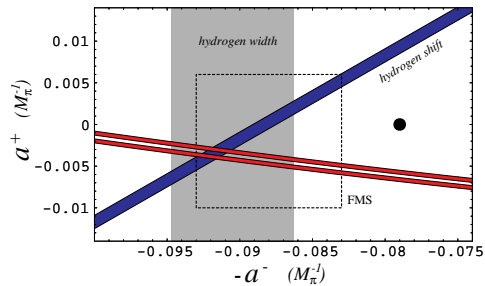}
  \end{center}
\vskip -0.5in
\caption{ 
$a^+$  vs $-a^-$~\protect\cite{BBEMP}. 
The light and dark shaded regions are from the experimental
pionic-hydrogen width and shift, respectively, 
taken from Ref.~\protect\cite{Schroder:uq}.
The dotted line encompasses the constraints from $\pi$-N phase shift 
data~\protect\cite{Fettes:1998ud}.
The dot is LO $\chi$PT. The two parallel bands are from
$\pi d$ scattering~\protect\cite{BBEMP} 
evaluated with the NLO wavefunction with a 
cutoff of $500$~MeV (upper curve) and $600$~MeV (lower curve).
 }
\label{fig:pid}
\vskip -0.1in
\end{figure}
Recently, W power-counting has been used by Beane {\it et al}
to examine $\pi d$ scattering in EFT~\cite{BBEMP}.
The EFT allows one to understand $\pi d$ scattering in the same framework as 
$\pi N$ scattering.  
Given the precise data from pionic hydrogen and from
pionic deuterium one can extract both the isoscalar and isovector $\pi N$
scattering lengths, $a^+$ and $a^-$ respectively, 
with high precision.
A plot of the constraints in the $a^+-a^-$ plane is shown in 
Fig.~\ref{fig:pid},
from which it is found that
$a^-=0.0918\pm 0.0013~m_\pi^{-1}$ and 
$a^+=-0.0034\pm 0.0007~m_\pi^{-1}$.
These values are in very good agreement with the recent analysis 
of Ref.~\cite{Ericson:2000md}.
A higher order calculation is required in order to further 
reduce the uncertainty in these values.

An alternate approach to EFT calculations has been advocated by 
Rho and collaborators. They argue that 
performing a chiral expansion of the current operators (or
whatever operator you are interested in) and using wavefunctions generated by 
the best modern potentials (``best'' being defined as those with lowest 
$\chi^2$ in the two- and three-nucleon sectors)
should be equivalent to the formal EFT expansion.
While I do not presently see how to formally justify this approach
it appears to numerically converge well, 
and its cut-off dependence is systematically reduced
as one performs calculations to higher orders.
With this method, Park {\it et al}~\cite{PMSVKRKMR} 
have computed the cross sections for $p\ p\rightarrow d\  e^+\ \nu_e$ and 
$p \ ^3{\rm He}\rightarrow ^4 {\rm He}\  e^+\  \nu_e$.
The wavefunctions are taken from the AV18/UIX potential while the current
operator is expanded to a given order in W counting.
To the order to which they work only one-body and two-body 
operators contribute.
The one unknown coefficient appearing in the two-body contribution, $L_{1,A}$,
is determined from the rate for tritium $\beta$-decay,
which they compute in the same framework.
While the coefficient of the two-body operator depends sensitively on the 
value of the cut-off, the complete two-body contribution to the rate is 
relatively insensitive to the
cut-off.  They find
$S_{\rm hep}(0) = (8.6\pm 1.3)\times 10^{-20}~{\rm keV-b}$.
This is a very encouraging result, as the calculation of 
$S_{\rm hep}(0)$ suffers from significant cancellations.
This method may provide a bridge between the formal EFT constructions
and the many-body methods developed in nuclear physics.

\section{The $m_q$-Dependence of the Two-Nucleon Sector}

\begin{figure}[!htb]
\begin{center}
\includegraphics*[width=0.4\linewidth,clip=true]{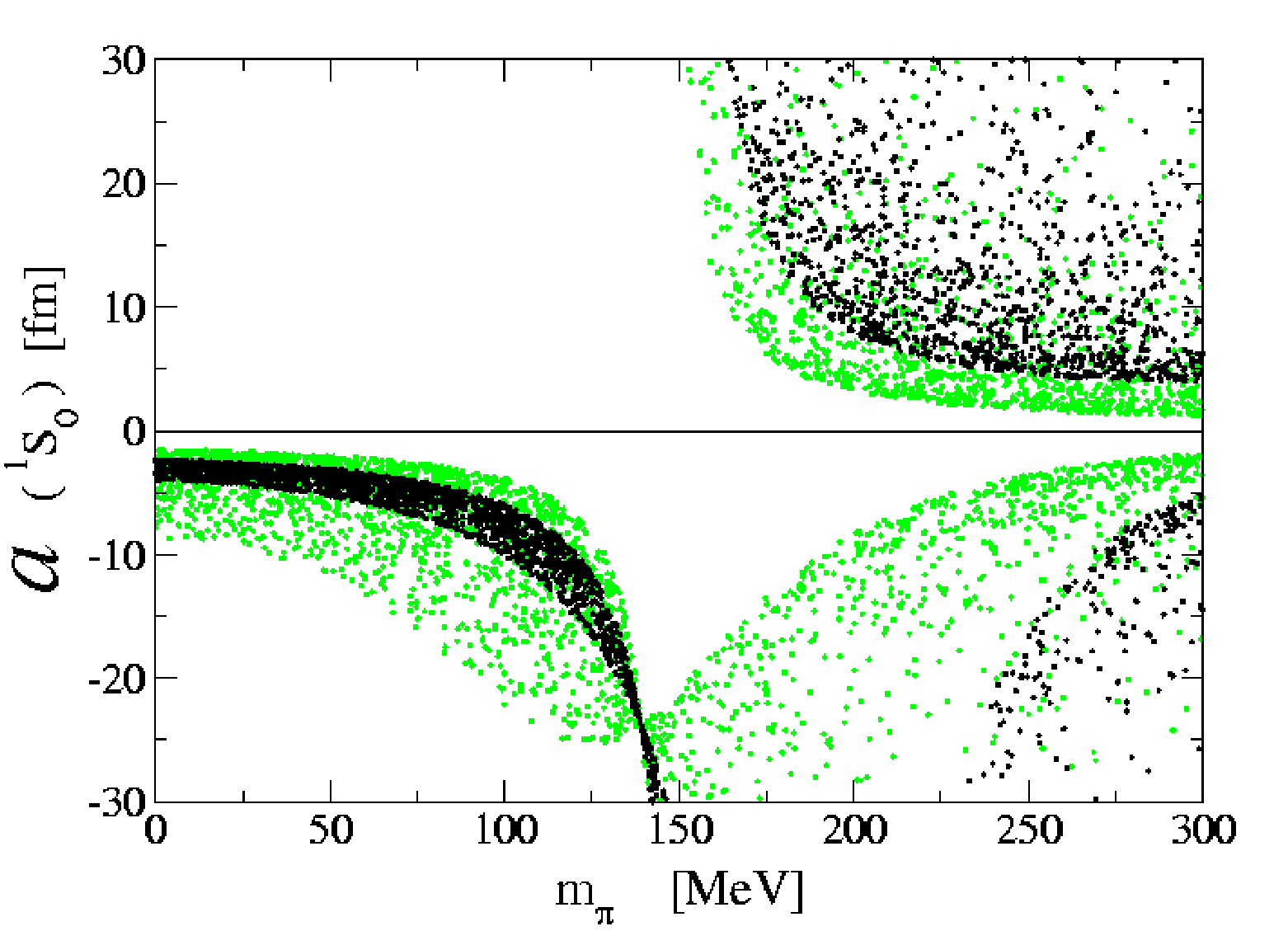}
\hskip 0.6in
\includegraphics*[width=0.4\linewidth,clip=true]{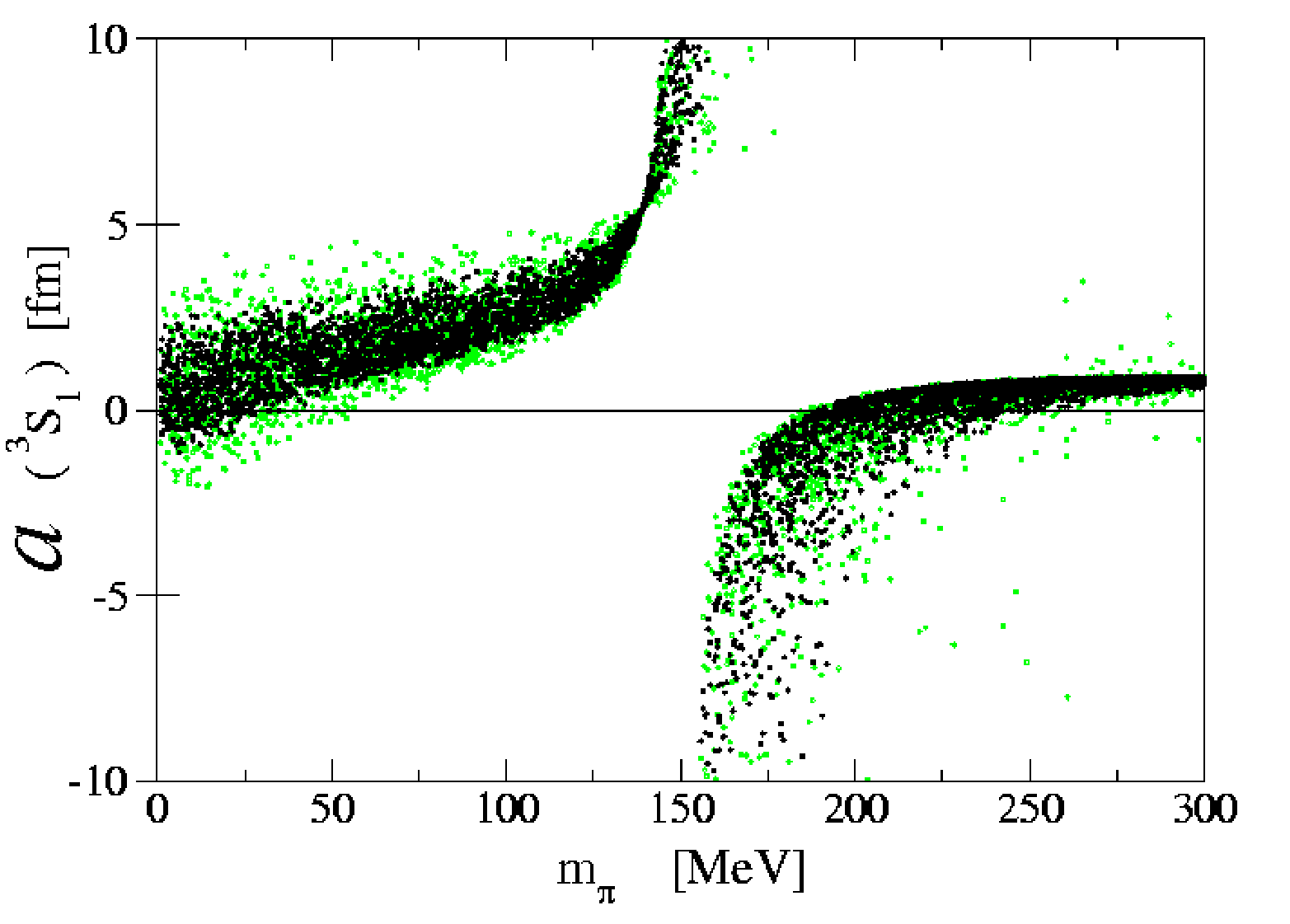}
\end{center}
\vskip -0.5in
\caption{
The left (right) panel shows
the scattering length in the $\si$-channel ($\siii$-channel)
as a function of the  pion mass.
The light gray region corresponds to $\eta=1/5$ and the 
black region corresponds to $\eta=1/15$.
In the $\siii$-channel
the parameter $\overline{d}_{16}$ is taken to be in the interval
$-2.61~{\rm GeV}^{-2} < \overline{d}_{16} < -0.17 ~{\rm GeV}^{-2}$
and $\overline{d}_{18}=-0.51~{\rm GeV}^{-2}$.
 }
\label{fig:1s0mq}
\vskip -0.1in
\end{figure}

In some recent papers by Beane and myself~\cite{BSmq} 
and also by Epelbaum, Glockle and Mei\ss ner~\cite{EMmq}
EFT was used to determine the $m_q$-dependence of scattering in 
the two-nucleon sector.
In the $\si$-channel KSW 
power-counting can be used to derive an analytic expression
for the scattering length,
\begin{eqnarray}
\frac{1}{a^{(\si)}}=\gamma +\frac{g_A^2 M_N}{8\pi f_\pi^2}\left[ m_\pi^2\,
  \log{\left({\mu\over m_\pi}\right)}+ (\gamma - m_\pi )^2
-(\gamma - \mu )^2 \right] 
-\frac{M_N m_\pi^2}{4\pi}(\gamma -\mu )^2 \ D_2  
\ \ ,
\end{eqnarray}
where $\gamma$ is a LO constant and $D_2 (\mu)$ is a combination 
of  coefficients of operators with a single insertion of $m_q$, that is
presently unknown.
The best that one can do at this point in time is to use naive dimensional
analysis (NDA) to estimate a range of reasonable values for $D_2$,
defined by a parameter $\eta \ll 1$~\cite{BSmq}.
The results of NDA are shown in Fig.~\ref{fig:1s0mq}.
NDA suggests that the di-neutron remains unbound in the 
chiral limit, while a relatively small increase in $m_q$ could lead to a 
bound di-neutron.

In the $\siii-\diii$ coupled channels the situation is somewhat more
complicated.
AT NLO in BBSvK counting not only does OPE contribute, but also the chiral
limit of two-pion exchange.
As a consequence, there are additional counterterms in the 
single nucleon sector
that contribute in this channel but do not contribute to 
the $\si$ channel, in particular $\overline{d}_{18}$ and 
$\overline{d}_{16}$ associated with the pion-nucleon interaction,
and $\overline{l}_4$ associated with $f_\pi$.
This is in addition to the $D_2 (\mu)$ contribution in the $\siii$ channel.
The allowed regions for $\overline{d}_{18}$ and 
$\overline{d}_{16}$ are given in Ref.\cite{Fettes:fd},
and $\overline{l}_4$ is known.
Fig.~\ref{fig:1s0mq} shows the presently allowed values of the scattering
length in the $\siii$ channel where we again have used NDA to estimate the 
possible values for $D_2$.
It is clear that for the range of parameters considered
the deuteron could be bound or unbound in the chiral limit,
and at present one cannot make a more definitive statement.
This last statement disagrees with the conclusion of Ref.~\cite{EMmq}.

\section{Conclusions}

There has been rapid progress in the
application of effective field theory to nuclear physics.
A consistent power-counting has been established,
improved model independent calculations of 
$NN\rightarrow NN$, $Nd\rightarrow Nd$, 
$\nu d\rightarrow \nu d$,
$p p\rightarrow d e \nu$, $p\  ^3{\rm He}\rightarrow ^4{\rm He}\  e \nu$,
$\gamma  d\rightarrow \gamma  d$ and $\pi  d\rightarrow \pi  d$ 
have been completed, and first efforts to describe halo-nuclei and hypernuclei
have been undertaken.
Further, efforts to understand fundamental questions 
about nuclear observables
have been made with interesting results.

There are still many goals to be attained.
It is clear that the calculations I have discussed here should 
be pushed to one higher order 
to convince ourselves about their precision.
Moreover, it is clear that parameters in 
the single nucleon sector need to be known to higher precision than they are
now in order  
to make precise statements in the multi-nucleon sectors.
Ultimately, a much closer link between the lattice QCD community and the EFT
community must be developed, as it is only with EFT that lattice calculations
of complex hadronic systems will be possible, and
further, it is only with lattice QCD that EFT will be able to make faithful
calculations in multi-nucleon systems.
This is clearly a symbiotic relationship that is yet to be realized.

\vskip 0.2in

I would like to thank the organizers of PaNIC02 for putting together a
stimulating meeting and inviting me to share the recent developments in EFT.
I would also like to thank those who have let me reproduce their work in these
proceedings.

\end{document}